\pdfoutput=1

\documentclass[preprint,5p]{elsarticle}

\usepackage{graphicx,amssymb}
\usepackage{epsf,psfig}
\usepackage{txfonts}
\usepackage{natbib}
\usepackage{hyperref}

\journal{International journal of Non-Linear Mechanics}

\begin{document}

\begin{frontmatter}
\title{Escape and collision dynamics in the planar equilateral restricted \\ four-body problem}

\author[]{Euaggelos E. Zotos\corref{cor}}
\ead{evzotos@physics.auth.gr}

\cortext[cor]{Corresponding author}

\address{Department of Physics, School of Science, \\
Aristotle University of Thessaloniki, \\
GR-541 24, Thessaloniki, \\
Greece}

\begin{abstract}
We consider the planar circular equilateral restricted four body-problem where a test particle of infinitesimal mass is moving under the gravitational attraction of three primary bodies which move on circular orbits around their common center of gravity, such that their configuration is always an equilateral triangle. The case where all three primaries have equal masses is numerically investigated. A thorough numerical analysis takes place in the configuration $(x,y)$ as well as in the $(x,C)$ space in which we classify initial conditions of orbits into four main categories: (i) bounded regular orbits, (ii) trapped chaotic orbits, (iii) escaping orbits and (iv) collision orbits. Interpreting the collision motion as leaking in the phase space we related our results to both chaotic scattering and the theory of leaking Hamiltonian systems. We successfully located the escape and the collision basins and we managed to correlate them with the corresponding escape and collision times of orbits. We hope our contribution to be useful for a further understanding of the escape and collision properties of motion in this interesting dynamical system.
\end{abstract}


\end{frontmatter}

\section{Introduction}
\label{intro}

Over the years several dynamical systems consisting of few bodies have been investigated and various models have been proposed in order to understand and explain the orbital behaviour of realistic celestial systems or as benchmark models where new mathematical theories can be tested. The most extensively studied dynamical system in celestial mechanics is, beyond any doubt, the classical restricted three-body problem, where the third body (test particle) is considered massless so as not to influence the motion of the two primaries which move in Keplerian orbits (circular or elliptical) around their common center of gravity \cite{dAT14,N04,N05,S67,Z15a,Z15b}. The modern applications to space mechanics and dynamics are probably even more cogent than the classical applications. Today numerous aspects in space dynamics are of paramount importance and of great interest. The applications of the restricted three-body problem create the basis of most of the lunar and planetary theories used for launching artificial satellites in the Earth-Moon system and in solar system in general.

In the same vein the restricted four-body problem is quite similar in the sense that the problem deals once more with the motion of an infinitesimal particle under the attraction of three primary bodies \cite{CRR64,CRB68,K10,MLJW08,SSD09a,SSD09b,VHW86}. There are many reasons justifying the study of the four-body problem (restricted or not). To begin with, there are many four-body systems in our Solar System which can be approximated, in a first order, by a four-body problem. A characteristic example is the Sun-Jupiter-Saturn system where the fourth body can be a planet, an asteroid or a satellite of Jupiter or Saturn. Another interesting example is the Sun-Earth-Moon system where the fourth body can be a space vehicle \cite{dAP05,J00,MPS07,SGJM95}. A special case of the Sun-Earth-Moon restricted four-body problem is the bi-circular problem, where the masses of the primary bodies are revolving in a quasi-bi-circular motion \cite{A02}.

The Sun-Jupiter-Trojan asteroid can also be viewed as a restricted four-body problem, where the primaries are in the particular configuration of an equilateral triangle. This special configuration is known as the planar equilateral restricted four-body problem (PERFBP). Many scientists studied this dynamical system \cite{CB10,MSML02,M06,RG06,SRB98a,SRB98b}. \cite{ARV09} investigated the PERFBP with equal masses, while \cite{BP11a} determined the total number of the equilibrium points for any value of the mass parameter and numerically explored their linear stability. They also computed some families of symmetric periodic orbits. Similar results were obtained in \cite{BGD13b} but for the case of two equal masses, while the existence of blue sky catastrophe around a specific collinear equilibrium point was presented in \cite{BGD13a}. In a recent paper \cite{BP13} a large number of families of non-symmetric periodic orbits around Jupiter and the Trojan asteroids was found. Moreover, in \cite{ARDV14} it was proved that any double collision can be regularized by using a Birkhoff-type transformations.

The restricted four-body problem has also applications in galactic dynamics. It is known that approximately two-thirds of the $10^{11}$ stars in our Galaxy belong to multi-stellar systems \cite{R04}. In particular, around one-fifth of these stars form triple systems, while a rough estimate suggests that a further one-fifth of these triple systems belongs to quadruple or higher systems, which can be modelled by the four-body problem \cite{ML99}.

A lot of work has been done regarding the equilibrium points of the PERFBP and their stability \cite {ASS15,BP11a,H80,L06,M81,PP13,P07,S78,SV15}. Another interesting issue is the location of periodic orbits in the PERFBP \cite{AB15,BP11b,BGD13b,SPB08}. In recent years many perturbing forces, such as the oblateness, radiation forces of the primaries, Coriolis and centrifugal force, variation of the masses of the primaries etc have been included in the study of PERFBP \cite{AES12,KAE07,KH08,KK13,KK14,PP13,SV15}.

In this paper we shall try to explore the orbital dynamics in the PERFBP by performing a systematic orbit classification using the numerical methods introduced in the pioneer works of \cite{N04} and \cite{N05}. The same numerical methods have also been successfully used in recent similar studies \cite{Z15a,Z15b,Z15c,RLS16}. The structure of the paper is as follows: In Section \ref{mod} we provide a detailed presentation of the principal aspects of the PERFBP. All the computational methods we used in order to determine the character of the orbits are described in Section \ref{cometh}. In the following Section, we conduct a thorough numerical investigation revealing the overall orbital structure (bounded regions and basins of escape/collision) of the system and how it is affected by the value of the Jacobi constant. Our paper ends with Section \ref{conc}, where the discussion and the conclusions of this work are given.

\section{Presentation of the mathematical model}
\label{mod}

Let us describe the basic properties of the PERFBP. We consider three primary bodies with masses $m_i$, $i = 1, 2, 3$, in a triangular (or Lagrangian) configuration in which the three primaries move in circular orbits in the same plane around their common center of mass. The three bodies are always located at the vertices of an equilateral triangle \cite{W41}. The fourth body is known as an infinitesimal mass (or test particle) and it moves in the same plane acting upon the attraction of the three primaries. It is assumed that the mass of the fourth body is so small that it's influence on the motion of the primaries is practically negligible.

We adopt a rotating rectangular system whose origin is the center of mass of the primaries which rotates with a uniform angular velocity, so that the centers of the three primaries to be fixed on the $(x,y)$-plane. Without loss of generality we assign the primary of mass $m_1$ on the positive $x$-axis at $C_1 = (x_1, 0)$. Then the other two primaries with masses $m_2$ and $m_3$, respectively are located at $C_2 = (x_2, \frac{1}{2})$ and $C_3 = (x_2, -\frac{1}{2})$, where $x_1 = \sqrt{3} \mu$, $x_2 = -\frac{\sqrt{3}}{2}(1 - 2\mu)$, while $\mu$ is the mass parameter. We normalize the units with the supposition that the sum of the masses and the distance between the primaries both be equal to unity. Therefore $m_1 = 1 - 2\mu$ and $m_2 = m_3 = \mu$, so that $m_1 + m_2 + m_3 = 1$.

Regarding the value of the mass parameter there are three limiting cases:
\begin{itemize}
  \item When $\mu = 0$ we obtain the rotating Kepler's central force problem with $m_1 = 1$ located at the origin of the coordinates.
  \item When $\mu = \frac{1}{2}$ we obtain the classical circular restricted three-body problem, with two equal masses $m_1 = m_2 =  \frac{1}{2}$, which is known as the Copenhagen problem.
  \item When $\mu = \frac{1}{3}$ we obtain the symmetric case with three primary bodies with masses equal to $\frac{1}{3}$.
\end{itemize}
In our study we shall consider the last case.

The forces experienced by the test particle in the coordinate system rotating with angular velocity $\omega = 1$ and origin at the center of the mass can be derived from the following total time-independent effective potential function
\begin{equation}
\Omega(x,y) = \frac{1 -  2\mu}{r_1} + \frac{\mu}{r_2} + \frac{\mu}{r_3} + \frac{1}{2}\left(x^2 + y^2 \right),
\label{pot}
\end{equation}
where
\[
r_1 = \sqrt{\left(x - x_1\right)^2 + y^2},
\]
\[
r_2 = \sqrt{\left(x - x_2\right)^2 + \left(y - \frac{1}{2}\right)^2},
\]
\begin{equation}
r_3 = \sqrt{\left(x - x_2\right)^2 + \left(y + \frac{1}{2}\right)^2},
\end{equation}
are the position vectors from the three primaries to the test particle, respectively. Using a synodical system we fixed the position of the primaries in order to eliminate the time-dependence in the potential function.

The scaled equations of motion describing the motion of the test body in the synodical coordinates $(x,y)$ read
\[
\Omega_x = \ddot{x} - 2\dot{y} = \frac{\partial \Omega(x,y)}{\partial x},
\]
\begin{equation}
\Omega_y = \ddot{y} + 2\dot{x} = \frac{\partial \Omega(x,y)}{\partial y},
\label{eqmot}
\end{equation}
where dots denote time derivatives, while the suffixes $x$ and $y$ indicate the partial derivatives of $\Omega(x,y)$ with resect to $x$ and $y$, respectively. Here it should be noted that Eqs. (\ref{eqmot}) are invariant under the symmetry
\begin{equation}
\Sigma: \left(t,x,y,\dot{x},\dot{y}\right) \rightarrow \left(-t,x,-y,-\dot{x},\dot{y}\right).
\label{sym}
\end{equation}

\begin{figure}[!tH]
\centering
\resizebox{\hsize}{!}{\includegraphics{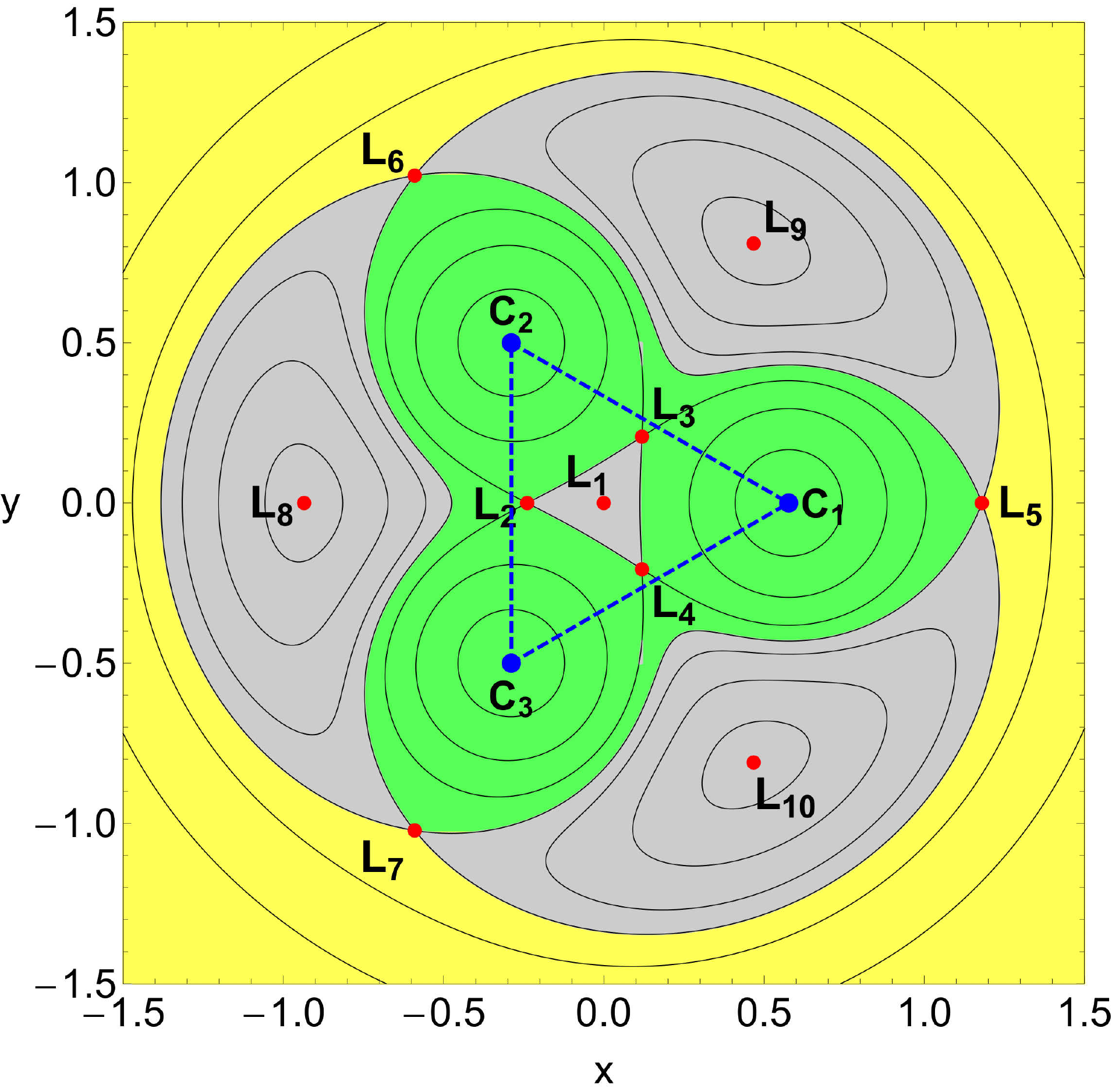}}
\caption{The isolines contours of the constant potential, the location of the centers of the three primaries (blue) and the position of the ten equilibrium points (red), for $\mu = \frac{1}{3}$. The interior region is indicated in green, the exterior region is shown in yellow, while the forbidden regions of motion are marked with grey. The equilateral triangle formed by the centers of the primaries is shown in blue dashed lines.}
\label{conts}
\end{figure}

The dynamical system (\ref{eqmot}) admits the well know Jacobi integral
\begin{equation}
J(x,y,\dot{x},\dot{y}) = 2\Omega(x,y) - \left(\dot{x}^2 + \dot{y}^2 \right) = C,
\label{ham}
\end{equation}
where $\dot{x}$ and $\dot{y}$ are the velocities, while $C$ is the Jacobi constant which is conserved and defines a three-dimensional invariant manifold in the total four-dimensional phase space. Thus, an orbit with a given value of it's energy integral is restricted in its motion to regions in which $C \leq 2\Omega(x,y)$, while all other regions are forbidden to the test body. If the problem is written in canonical coordinates, then the Jacobi integral corresponds to the value of the Hamiltonian and it is known as the total orbital energy. The value of the total orbital energy $E$ is related with the Jacobi constant by $C = - 2E$. It should be emphasized that the existence of the Jacobi integral, allows us to study the problem by fixing the energy level of the value of the Jacobi constant.

\begin{figure*}[!tH]
\centering
\resizebox{\hsize}{!}{\includegraphics{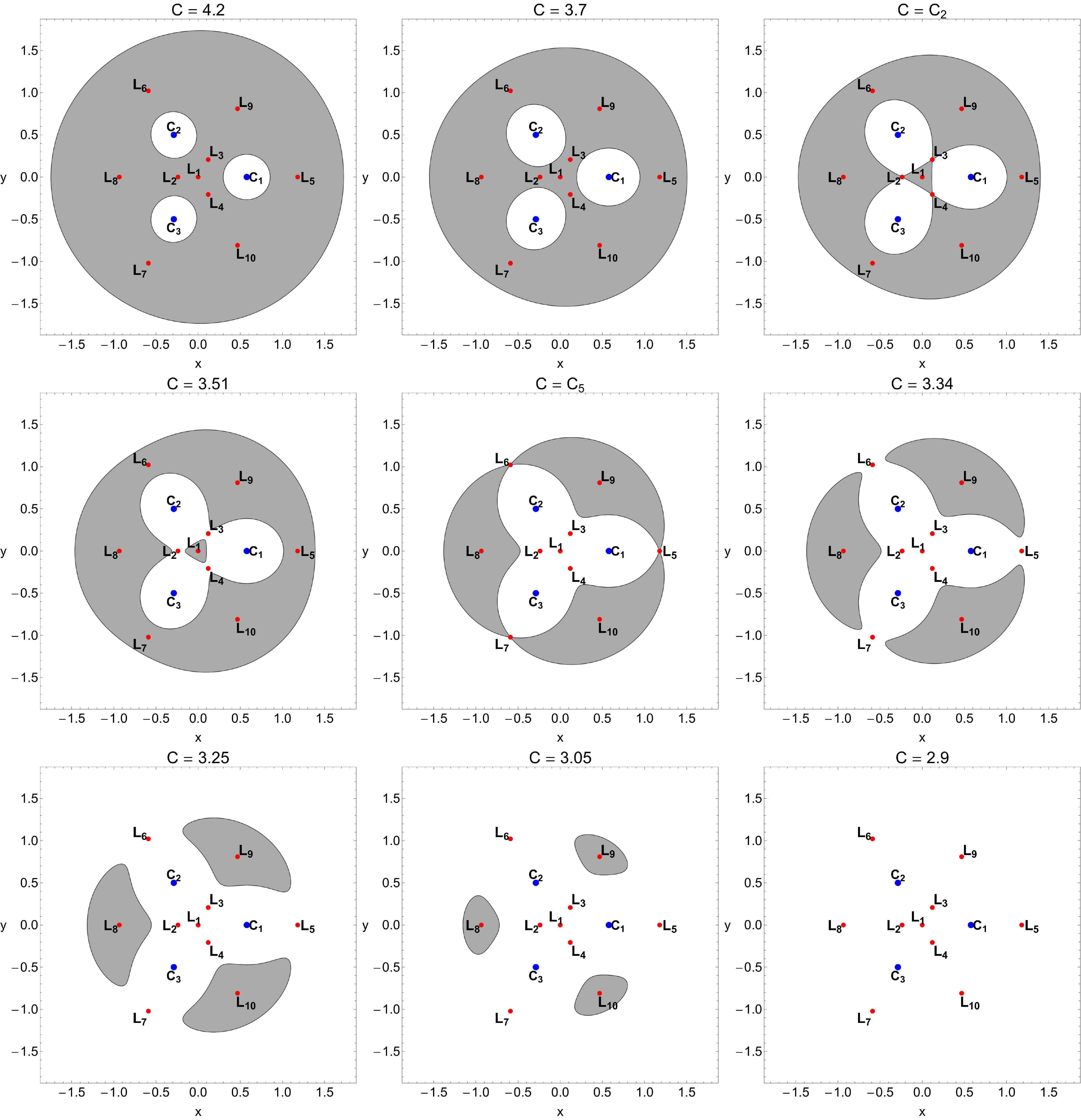}}
\caption{Evolution of the structure of the Hill's regions configurations for the PERFBP. The white domains correspond to the Hill's regions, gray shaded domains indicate the forbidden regions, while the thick black lines depict the Zero Velocity Curves (ZVCs). The red dots pinpoint the position of the equilibrium points, while the positions of the centers of the three primary bodies are indicated by blue dots.}
\label{HRCs}
\end{figure*}

In the classical restricted three-body problem there are five coplanar equilibrium points \cite{S67}. In the PERFBP on the other hand, \cite{L06} proved that the existence as well as the total number of the equilibrium points (collinear and non-collinear) strongly depends on the value of the mass parameter (see also \cite{BP11a}). In our case where all primaries have the same mass $m_1 = m_2 = m_3 = \frac{1}{3}$, the system admits four collinear (on the $x$-axis) equilibrium points and six non-collinear (off the $x$-axis) ones. Due to the equality of the masses of the primaries the PERFBP admits a symmetry and the ten equilibrium points lie on the $(x,y)$-plane symmetrically to the axes of symmetry $y = 0$, $y = \sqrt{3}$ and $y = -\sqrt{3}$. Fig. \ref{conts} shows the position of the ten equilibrium points along with the centers of the primary bodies, while in Table \ref{table1} we provide the exact coordinates of the equilibrium points. All ten equilibrium points are unstable \cite{ARV09,BP11a}. A thorough discussion of the equilibrium points can be found in \cite{ARD03}, \cite{M87} and \cite{S78}. Furthermore, an analytical examination of the stability of the equilibrium points can be found in \cite{BP12}, while a numerically in \cite{BP11a}.

\begin{table*}[!ht]
\begin{center}
   \caption{The position of the ten equilibrium points and the corresponding critical values of the Jacobi constant when $\mu = \frac{1}{3}$.}
   \label{table1}
   \setlength{\tabcolsep}{2pt}
   \begin{tabular}{@{}lcc}
      \hline
      $L_i$ & $(x,y)$ & $C_i$ \\
      \hline
      $L_1$ &                                    (0,0) & 3.464101615137753 \\
      $L_2$ &                  (-0.238958309195350, 0) & 3.527366587689146 \\
      $L_3$ &   (0.119479154597674, 0.206943966208549) & 3.527366587689146 \\
      $L_4$ &  (0.119479154597674, -0.206943966208549) & 3.527366587689146 \\
      $L_5$ &                   (1.179998404889433, 0) & 3.358035160821999 \\
      $L_6$ &  (-0.589999202444716, 1.021908595059364) & 3.358035160821999 \\
      $L_7$ & (-0.589999202444716, -1.021908595059364) & 3.358035160821999 \\
      $L_8$ &                  (-0.935185966672243, 0) & 2.946725190762343 \\
      $L_9$ &   (0.467592983336121, 0.809894804400872) & 2.946725190762343 \\
   $L_{10}$ &  (0.467592983336121, -0.809894804400872) & 2.946725190762343 \\
      \hline
   \end{tabular}
\end{center}
\end{table*}

Using the surface $2\Omega(x,y) = C$ we can determine the regions on the configuration $(x,y)$-plane in which the test particle is allowed to move for a given value of the Jacobi constant $C$. The projection of this surface on the $(x,y)$-plane determine the Zero Velocity Curves (ZVCs) of the PERFBP. \cite{ARV09} and \cite{S78} investigated the ZVCs for specific values of the mass parameter. The Jacobi constant values at the equilibrium points $L_i, i = 1, ..., 10$ are denoted by $C_i$ and are critical values. It should be noted that due to the symmetry of the PERFBP with equal masses $(m_1 = m_2 = m_3)$ and the existence of the three axes of symmetry we have that $C_2 = C_3 = C_4$, $C_5 = C_6 = C_7$ and $C_8 = C_9 = C_{10}$. In Table \ref{table1} we provide the critical values of the Jacobi constant.

The projection of the four-dimensional phase space onto the configuration (or position) space $(x,y)$ is called the Hill's regions and is divided into three domains: (i) the interior region, (ii) the exterior region and (iii) the forbidden regions (see Fig. \ref{conts}). The boundaries of these Hill's regions are the ZVCs because they are the locus in the configuration $(x,y)$ space where the kinetic energy vanishes. The structure of the Hill's regions strongly depends on the value of the Jacobi constant. There are four distinct cases regarding the Hill's regions:
\begin{itemize}
  \item $C > C_2$: The test particle is allowed to move either very close to each primary or far away from them without being able to travel between the primaries.
  \item $C_5 < C < C_2$: The channels between the primaries open therefore the fourth body can freely move from one allowed region to another allowed region. However, the test particle still cannot enter the exterior region thus fending off the primaries.
  \item $C_8 < C < C_5$: Escape channels emerge and the fourth body starting from the interior region is allowed to enter the exterior region and escape to infinity.
  \item $C < C_8$: The forbidden regions disappear, so motion over the entire configuration $(x,y)$ space is possible.
\end{itemize}
In Fig. \ref{HRCs}(a-h) we present the evolution of the structure of the Hill's region configurations for several values of the Jacobi constant. It is evident that as the value of the Jacobi constant decreases several doorways appear through which the test particle can enter the several allowed region of motion. For $C < C_5$ we observe the presence of the three openings (exit channels) at the equilibrium points $L_5$, $L_6$ and $L_7$ through which the test particle can enter the exterior region and then leak out. In fact, we may say that these two exits act as hoses connecting the interior region of the system with the ``outside world" of the exterior region.

\section{Computational methods and criteria}
\label{cometh}

The motion of the fourth body is restricted to a three-dimensional surface $C = const$, due to the existence of the Jacobi integral. With polar coordinates $(r,\phi)$ in the center of the mass system of the corotating frame the condition $\dot{r} = 0$ defines a two-dimensional surface of section, with two disjoint parts $\dot{\phi} < 0$ and $\dot{\phi} > 0$. Each of these two parts has a unique projection onto the configuration $(x,y)$ space. In order to explore the orbital structure of the system we need to define samples of initial conditions of orbits whose properties will be identified. For this purpose, we define for several values of the Jacobi constant $C$ dense uniform grids of $1024 \times 1024$ initial conditions regularly distributed on the configuration $(x,y)$ space inside the area allowed by the value of the Jacobi constant. Following a typical approach, the orbits are launched with initial conditions inside a certain region, called scattering region, which in our case is a square grid with $-2\leq x,y \leq 2$.

In the PERFBP the configuration space extends to infinity thus making the identification of the type of motion of the test particle for specific initial conditions a rather demanding task. There are three possible types of motion for the fourth body: (i) bounded motion around one of the primaries, or even around all of them; (ii) escape to infinity; (iii) collision into one of the three primaries. Now we need to define appropriate numerical criteria for distinguishing between these three types of motion. The motion is considered as bounded if the test body stays confined for integration time $t_{\rm max}$ inside the system's disk with radius $R_d$ and center coinciding with the center of mass origin at $(0,0)$. Obviously, the higher the values of $t_{\rm max}$ and $R_d$ the more plausible becomes the definition of bounded motion and in the limit $t_{\rm max} \rightarrow \infty$ the definition is the precise description of bounded motion in a finite disk of radius $R_d$. Consequently, the higher these two values, the longer the numerical integration of initial conditions of orbits lasts. In our calculations we choose $t_{\rm max} = 10^4$ and $R_d = 10$ as in \cite{N04,N05} and \cite{Z15a,Z15b}. We decided to include a relatively high disk radius $(R_d = 10)$ in order to be sure that the orbits will certainly escape from the system and not return back to the interior region. Furthermore, it should be emphasized that for low values of $t_{\rm max}$ the fractal boundaries of stability islands corresponding to bounded motion become more smooth. Moreover, an orbit is identified as escaping and the numerical integration stops if the fourth body intersects the system's disk with velocity pointing outwards at a time $t_{\rm esc} < t_{\rm max}$. Finally, a collision with one of the primaries occurs if the fourth body, assuming it is a point mass, crosses the disk with radius $R_m$ around the primary. For all primaries we choose $R_{m_1} = R_{m_2} = R_{m_3} = 10^{-4}$. In \cite{N04,N05} it was shown that the radii of the primaries influence the area of collision and escape basins.

As it was stated earlier, in our computations, we set $10^4$ time units as a maximum time of numerical integration. The vast majority of escaping orbits (regular and chaotic) however, need considerable less time to escape from the system (obviously, the numerical integration is effectively ended when an orbit moves outside the system's disk and escapes). Nevertheless, we decided to use such a vast integration time just to be sure that all orbits have enough time in order to escape. Here we should clarify, that orbits which do not escape after a numerical integration of $10^4$ time units are considered as non-escaping or trapped.

The phase space is divided into the escaping, non-escaping and collision space. Usually, the vast majority of the non-escaping space is occupied by initial conditions of regular orbits forming stability islands where a third integral is present. In many dynamical systems however, trapped chaotic orbits have also been observed \cite{Z15d}. Therefore, we decided to distinguish between regular non-escaping orbits and trapped chaotic motion. Over the years, several chaos indicators have been developed in order to safely distinguish between ordered and chaotic orbits. In our case, we choose to use the Smaller ALingment Index (SALI) method \cite{S01}. The SALI has been proved a very fast, reliable and effective tool \cite{ZC13,Z14}. The determination of the nature of an orbit is obtained from the final value of the SALI at the end of the numerical integration. In particular, if SALI $> 10^{-4}$ the motion is regular, while if SALI $< 10^{-8}$ the motion is chaotic. If the final value of SALI lies in the interval $[10^{-4}, 10^{-8}]$ then we have the case of a ``sticky" orbit\footnote{A stick orbit behaves for a long time interval as a regular one before revealing its true chaotic nature.} and further numerical integration is required for obtaining the true nature of the orbit.

All calculations reported in this paper were performed using a double precision Bulirsch-Stoer \verb!FORTRAN 77! algorithm \cite{PTVF92} with a variable time step. Here we should emphasize, that our previous numerical experience suggests that the Bulirsch-Stoer integrator is both faster and more accurate than a double precision Runge-Kutta-Fehlberg algorithm of order 7 with Cash-Karp coefficients. Throughout all our computations, the Jacobian energy integral (Eq. (\ref{ham})) was conserved better than one part in $10^{-11}$, although for most orbits it was better than one part in $10^{-12}$. For collision orbits where the fourth body moves inside a region of radius $10^{-2}$ around one of the primaries the Lemaitre's global regularization method is applied. All graphics presented in this work have been created using Mathematica$^{\circledR}$ \cite{W03}.

\section{Numerical results \& Orbit classification}
\label{numres}

In this section we shall classify initial conditions of orbits in the $\dot{\phi} < 0$ part\footnote{We choose the $\dot{\phi} < 0$ instead of the $\dot{\phi} > 0$ part simply because in \cite{Z15a} we seen that it contains more interesting orbital content.} of the surface of section $\dot{r} = 0$ into four main categories: (i) bounded regular orbits; (ii) trapped chaotic orbits; (iii) escaping orbits and (iv) collision orbits. Moreover, two additional properties of the orbits will be examined: (i) the time-scale of the collision and (ii) the time-scale of the escape (we shall also use the terms escape period or escape rate). Our main numerical task will be to explore these dynamical quantities for various values of the total orbital energy, or in other words for various values of the Jacobi constant. In particular, we will consider four different cases which correspond to the four possible Hill's regions configurations.

Looking at Fig. \ref{conts} it becomes evident that the PERFBP admits a $2\pi/3$ symmetry. In order to maintain this symmetry to the color-coded grids on the configuration $(x,y)$ space we should use initial conditions expressed in polar coordinates, rather than in Cartesian coordinates. The approach of polar coordinates has been used for the H\'{e}non-Heiles system which also admits the same type of symmetry \citep[e.g.,][]{AVS01,AVS09,SS10}.

At this point we would like to clarify how the initial conditions of orbits are generated in the $\dot{\phi} < 0$ part of the surface of section $\dot{r} = 0$. The conditions $\dot{\phi} < 0$ and $\dot{r} = 0$ in polar coordinates along with the existence of the Jacobi integral of motion (\ref{ham}) suggest that the four initial conditions of orbits in cartesian coordinates are
\begin{eqnarray}
x &=& x_0, \ \ \ y = y_0, \nonumber\\
\dot{x_0} &=& \frac{y_0}{r}\sqrt{2\Omega(x,y) - C}, \nonumber\\
\dot{y_0} &=& - \frac{x_0}{r}\sqrt{2\Omega(x,y) - C},
\end{eqnarray}
where $r = \sqrt{x_0^2 + y_0^2}$. Thus the initial conditions of orbits on the $(x,y)$-plane are classified into bounded orbits, unbounded or escaping orbits and collisional orbits. In this special type of Poincar\'{e} Surface of Section (PSS) the phase space emerges as a close and compact mix of escape basins, collisional basins and stability regions. Our numerical calculations indicate that apart from the escaping and collisional orbits there is also a considerable amount of non-escaping orbits. In general terms, the majority of non-escaping regions corresponds to initial conditions of regular orbits, where an adelphic integral of motion is present, restricting their accessible phase space and therefore hinders their escape.

In the following the orbit classification in the configuration space of the PERFBP will be conducted using color-coded grids or orbit type diagrams (OTDs) thus following the methods introduced in \cite{N04} and \cite{N05}. In these diagrams, a specific color is assigned to each pixel according to the type of the particular orbit. We can say that these color-coded plots are in fact a modern version of the classical Poincar\'{e} Surface of Section (PSS) where the phase space emerges as a compact mix of bounded, escape and collision basins.

\subsection{Case I: $C > C_2$}
\label{sb1}

\begin{figure*}[!tH]
\centering
\resizebox{\hsize}{!}{\includegraphics{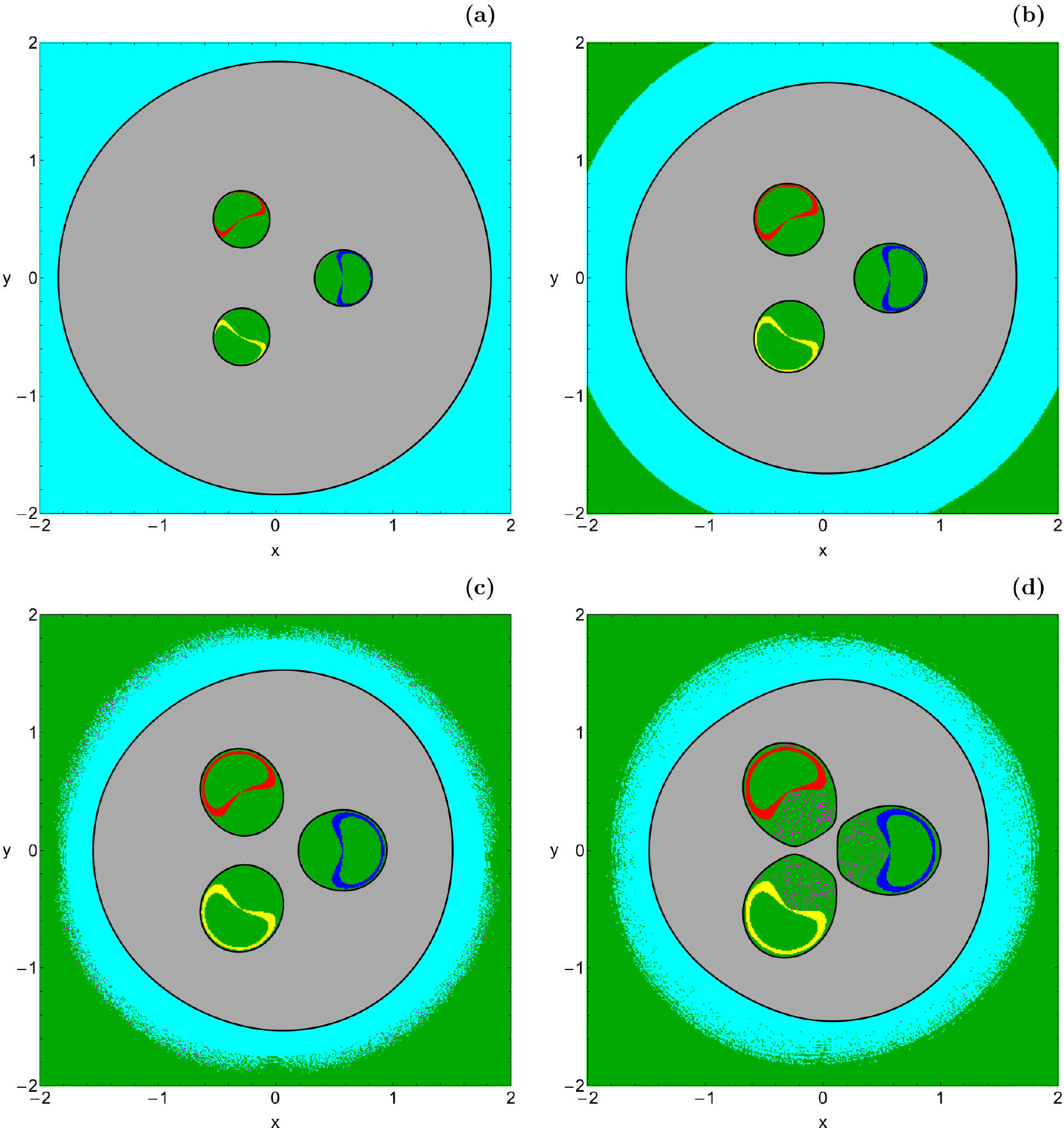}}
\caption{The orbital structure of the $\dot{\phi} < 0$ part of the surface of section $\dot{r} = 0$ when (a-upper left): $C = 4.50$; (b-upper right): $C = 4.00$; (c-lower left): $C = 3.70$; (d-lower right): $C = 3.54$. The color code is the following: bounded regular orbits (green), trapped chaotic orbits (magenta), collisional orbits to primary 1 (blue), collisional orbits to primary 2 (red), collisional orbits to primary 3 (yellow), and escaping orbits (cyan).}
\label{HR1}
\end{figure*}

\begin{figure*}[!tH]
\centering
\resizebox{\hsize}{!}{\includegraphics{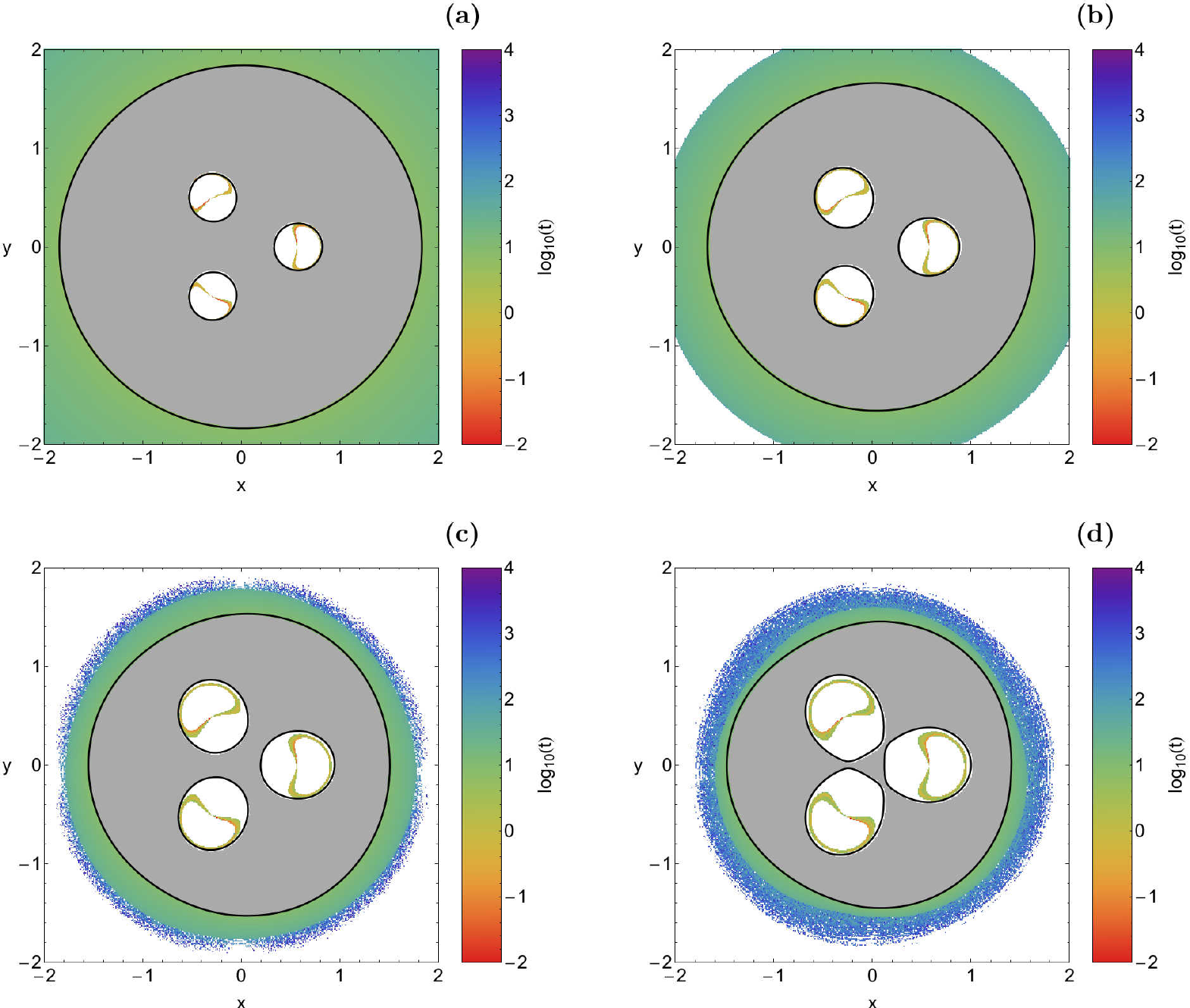}}
\caption{Distribution of the escape and collisional time of the orbits on the $\dot{\phi} < 0$ part of the surface of section $\dot{r} = 0$ for the values of the Jacobi constant of Fig. \ref{HR1}(a-d). The darker the color, the larger the escape/collisional time. Initial conditions of bounded regular orbits and trapped chaotic orbits are shown in white.}
\label{HR1t}
\end{figure*}

Our first case under investigation concerns the scenario where the Hill's regions configurations consist of small disks around the primaries together with an unbounded component having as a boundary a closed curve around the primary bodies. In Fig. \ref{HR1}(a-d) the OTD decompositions of the $\dot{\phi} < 0$ part of the surface of section $\dot{r} = 0$ reveal the orbital structure of the configuration $(x,y)$ space for four values of the Jacobi constant $C$. The black solid lines denote the ZVC, while the inaccessible forbidden regions are marked in gray. The color of a point represents the orbit type of the fourth body which has been launched with pericenter position at $(x,y)$. When $C = 4.5$ we see in Fig. \ref{HR1}a that inside the disks around the primaries there are, as expected, only two types of initial conditions of orbits: (i) bounded regular orbits and (ii) collisional orbits. In particular the initial conditions of collisional orbits form thin layers inside the stability islands. Due to the symmetry of the PERFBP with three equal masses the orbital structure around each primary is the same. The exterior region is dominated by initial conditions of orbits that escape from the system. In Fig. \ref{HR1}b where $C = 4.0$ we observe that the orbital structure of the interior regions remains unperturbed; only the size of the disks increases. In the exterior region on the other hand, we identify four parts of bounded orbits. Additional numerical calculations, not shown here, reveal that the four parts are actually pieces of an annulus of initial conditions of bounded regular orbits. These regular orbits however circulate around all three primaries. The ring-shaped stability island inside the escape region is also present for $C = 4.5$, but it is located in larger radius from the mass center at $(0,0)$ and therefore it is not visible in Fig. \ref{HR1}a. Things are quite similar in Fig. \ref{HR1}c where we present the orbital structure of the configuration $(x,y)$-plane for $C = 3.7$. However with a much closer look it is seen that the smoothness of the boundary in the exterior region between the basin of escape and the stability annulus is destroyed. The value of the Jacobi constant is further decreased and in Fig. \ref{HR1}d for $C = 3.54$ an interesting phenomenon takes place. Inside the disks around the primaries one can easily identify delocalized initial conditions of chaotic orbits. These initial conditions of chaotic orbits do not form any chaotic layer but they are randomly scattered inside the stability regions.

In the following Fig. \ref{HR1t}(a-d) we show how the escape and collisional times of orbits are distributed on the configuration $(x,y)$ space for the four values of the Jacobi constant discussed in Fig. \ref{HR1}(a-d). Light reddish colors correspond to fast escaping/collisional orbits, dark blue/purple colors indicate large escape/collisional times, while white color denote stability islands of regular motion and trapped chaotic motion. Note that the scale on the color bar is logarithmic. Inspecting the spatial distribution of various different ranges of escape time, we are able to associate medium escape time with the stable manifold of a non-attracting chaotic invariant set, which is spread out throughout this region of the chaotic sea, while the largest escape time values on the other hand, are linked with sticky motion around the stability islands of the two primary bodies. As for the collisional time we see that a small portion of orbits with initial conditions very close to the vicinity of the centers of the primaries collide with them almost immediately, within the first time steps of the numerical integration. Looking more carefully at Fig. \ref{HR1t}(a-d) we clearly observe that as the value of the Jacobi constant decreases the escape time of orbits which form the basin of escape in the exterior region gradually increases. This is true especially for initial conditions of orbits very close to the boundary between the basin of escape and the stability annulus.

\subsection{Case II: $C_5 < C < C_2$}
\label{sb2}

\begin{figure*}[!tH]
\centering
\resizebox{\hsize}{!}{\includegraphics{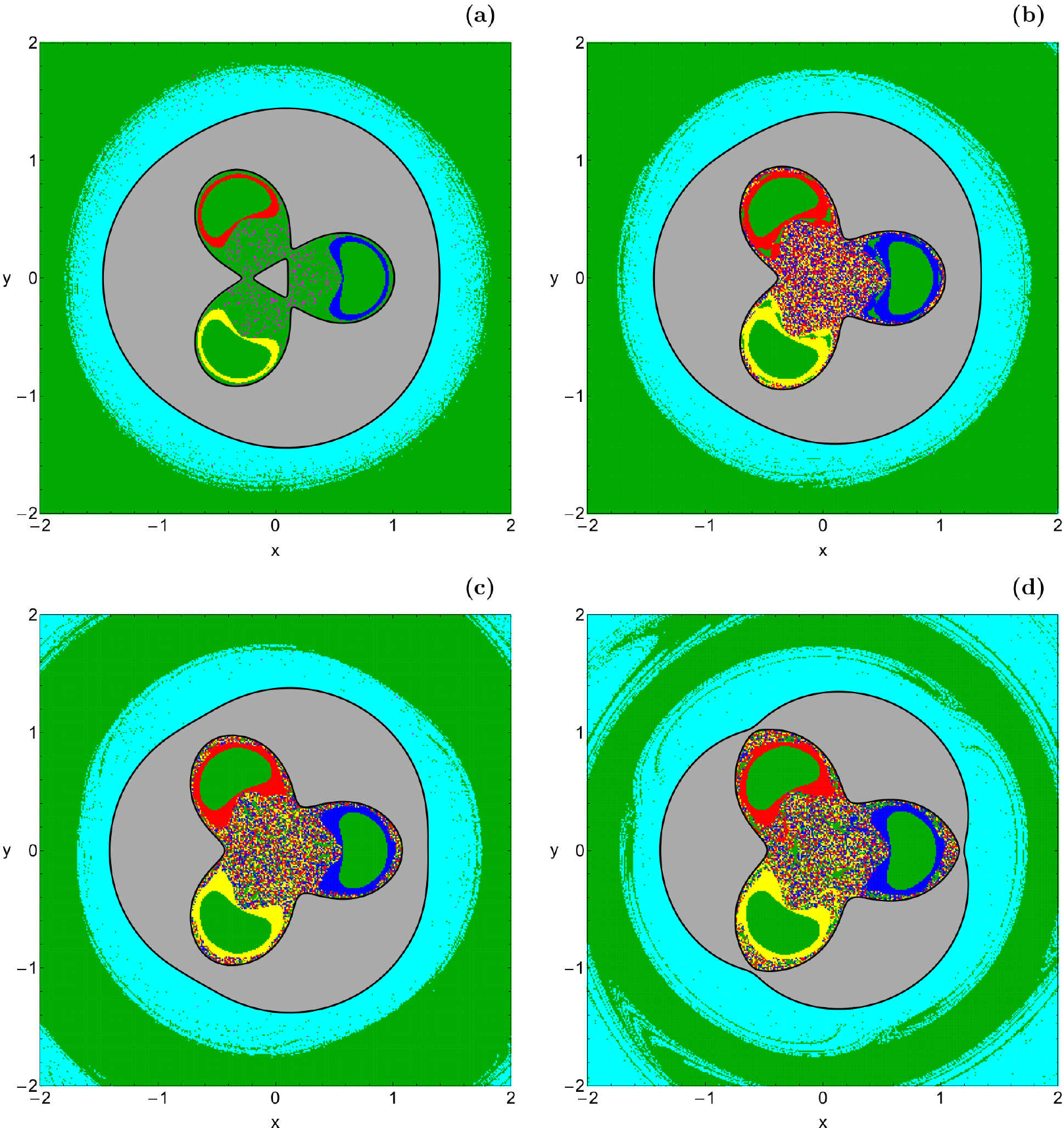}}
\caption{The orbital structure of the $\dot{\phi} < 0$ part of the surface of section $\dot{r} = 0$ when (a-upper left): $C = 3.52$; (b-upper right): $C = 3.46$; (c-lower left): $C = 3.41$; (d-lower right): $C = 3.36$. The color code is the same as in Fig. \ref{HR1}.}
\label{HR2}
\end{figure*}

\begin{figure*}[!tH]
\centering
\resizebox{\hsize}{!}{\includegraphics{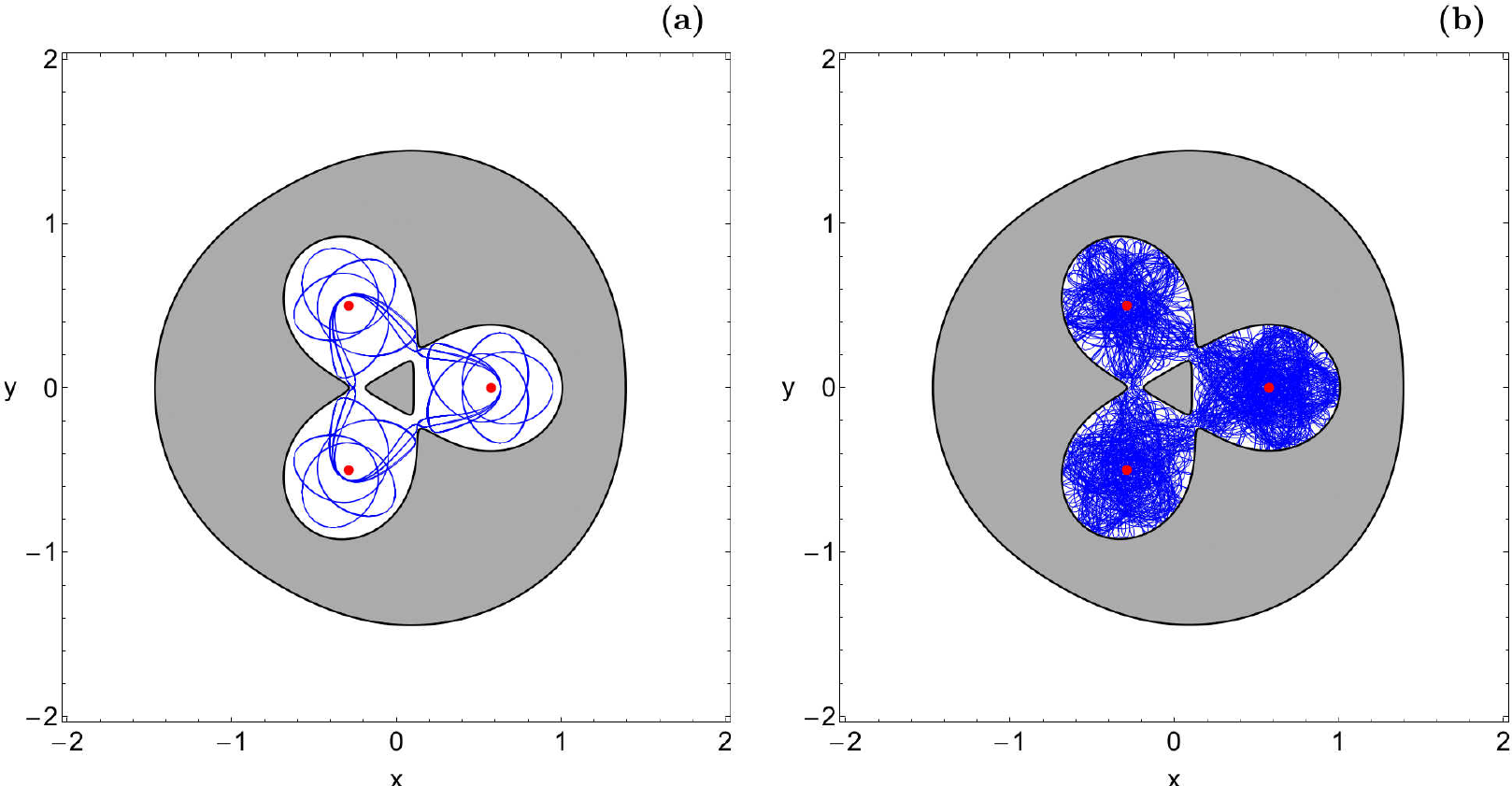}}
\caption{A characteristic example of (a-left): a quasi-periodic orbit with $(x_0, y_0) = (0.20689655, 0.05835544)$ and (b-right): a trapped chaotic orbit with $(x_0, y_0) = (0.20689655, 0.11405841)$, when $C = 3.52$. Both orbits were integrated for 500 time units and they circulate around all three primary bodies. The gray shaded domains indicate the forbidden regions, while the red dots denote the positions of the centers of the three primaries.}
\label{orbs}
\end{figure*}

\begin{figure*}[!tH]
\centering
\resizebox{\hsize}{!}{\includegraphics{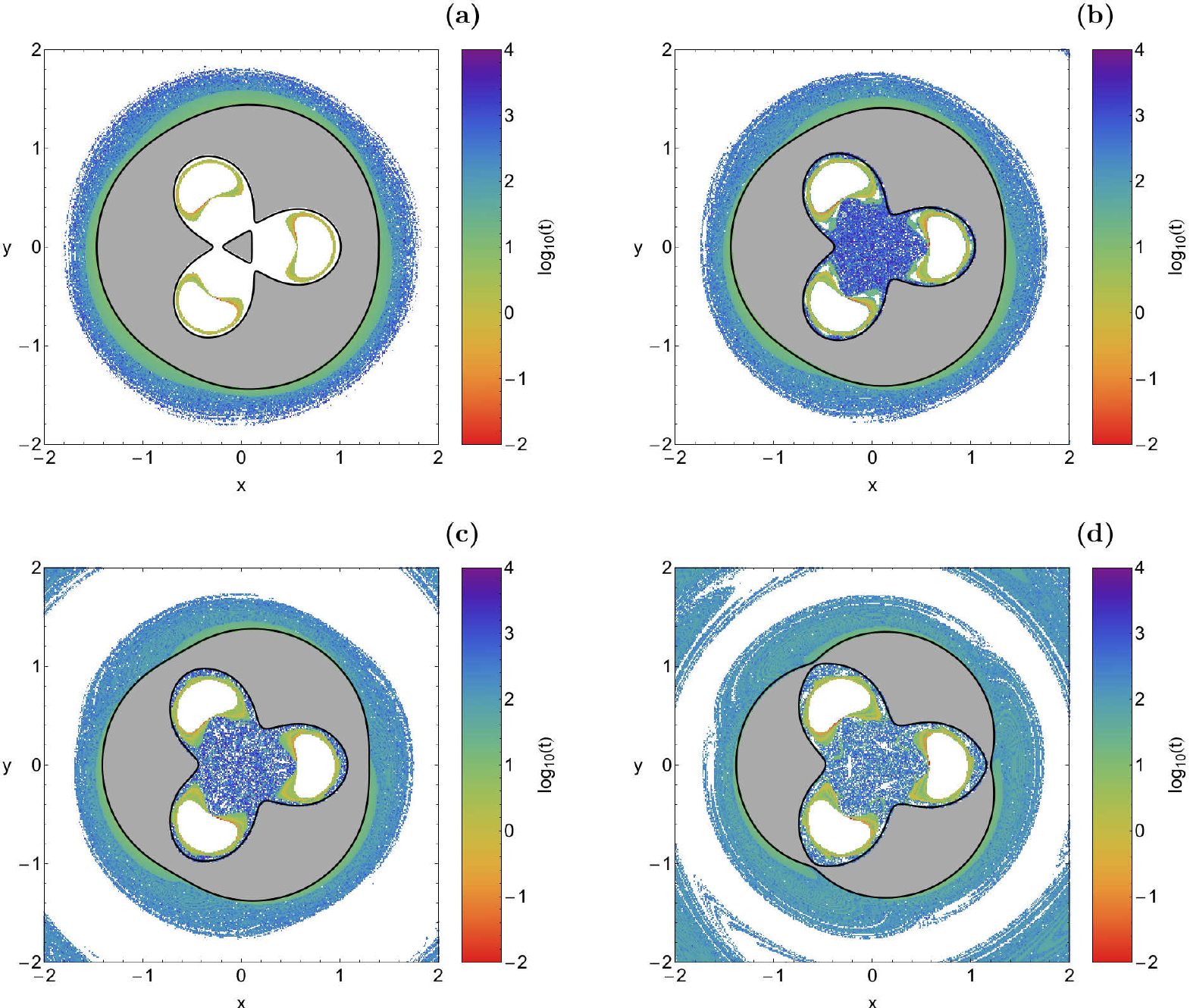}}
\caption{Distribution of the escape and collisional time of the orbits on the $\dot{\phi} < 0$ part of the surface of section $\dot{r} = 0$ for the values of the Jacobi constant of Fig. \ref{HR2}(a-d).}
\label{HR2t}
\end{figure*}

We continue our exploration considering the second Hill's regions configurations in which the test particle is allowed to move around all three primaries in the interior region. The orbital structure of the configuration $(x,y)$ space is unveiled in Fig. \ref{HR2}(a-d) through the OTD decompositions of the $\dot{\phi} < 0$ part of the surface of section $\dot{r} = 0$. The pattern in Fig. \ref{HR2}a for $C = 3.52$ is very similar to that discussed earlier in Fig. \ref{HR1}d with the main difference that now the transport channels between the primaries are open. Initial conditions of trapped chaotic orbits in the interior region are still present. In Fig. \ref{orbs}a we present an example of a quasi-periodic orbit and a trapped chaotic orbit which both of them circulate around all three primaries. The orbital structure of the central interior region changes drastically in Fig. \ref{HR2}b for $C = 3.46$, where the small forbidden region around the center disappears. We observe that the central region is highly fractal and it is composed of a rich mixture of initial conditions of collisional, bounded, and trapped chaotic orbits. When we state that an area is fractal we mean that it has a fractal-like geometry. The orbital content of the interior region remains the same in Fig. \ref{HR2}c for $C = 3.41$. However, it is seen that in the exterior region the size of the stability annulus is reduced and its upper boundaries are now visible. For $C = 3.36$ we see in Fig. \ref{HR2}d that the reduction of the size of the stability annulus continues. Moreover, since the boundaries of the stability annulus and the escape basins are somehow indistinct, we may say that the outer stability islands starts to destabilizes. For all the values of the Jacobi constant in the second Hill's regions configurations the collision basins around the stability islands of the three primaries are present and as the value of $C$ decreases the size of the collisional basins seems to increases.

The distribution of the escape and collisional times of orbits on the configuration $(x,y)$ space is shown in Fig. \ref{HR2t}(a-d). One may observe that the results are very similar to those presented earlier in Fig. \ref{HR1t}(a-d), where we found that orbits with initial conditions inside the escape and collisional basins have the smallest escape/collision rates, while on the other hand, the longest escape/collisional rates correspond to orbits with initial conditions in the fractal regions of the OTDs. We see that the collision time of orbits with initial conditions in the central fractal region is reduced as we proceed to smaller values of $C$. In the same vein, the escape rate of orbits in the exterior region is also reduced with decreasing value of the Jacobi constant.

\subsection{Case III: $C_8 < C < C_5$}
\label{sb3}

\begin{figure*}[!tH]
\centering
\resizebox{\hsize}{!}{\includegraphics{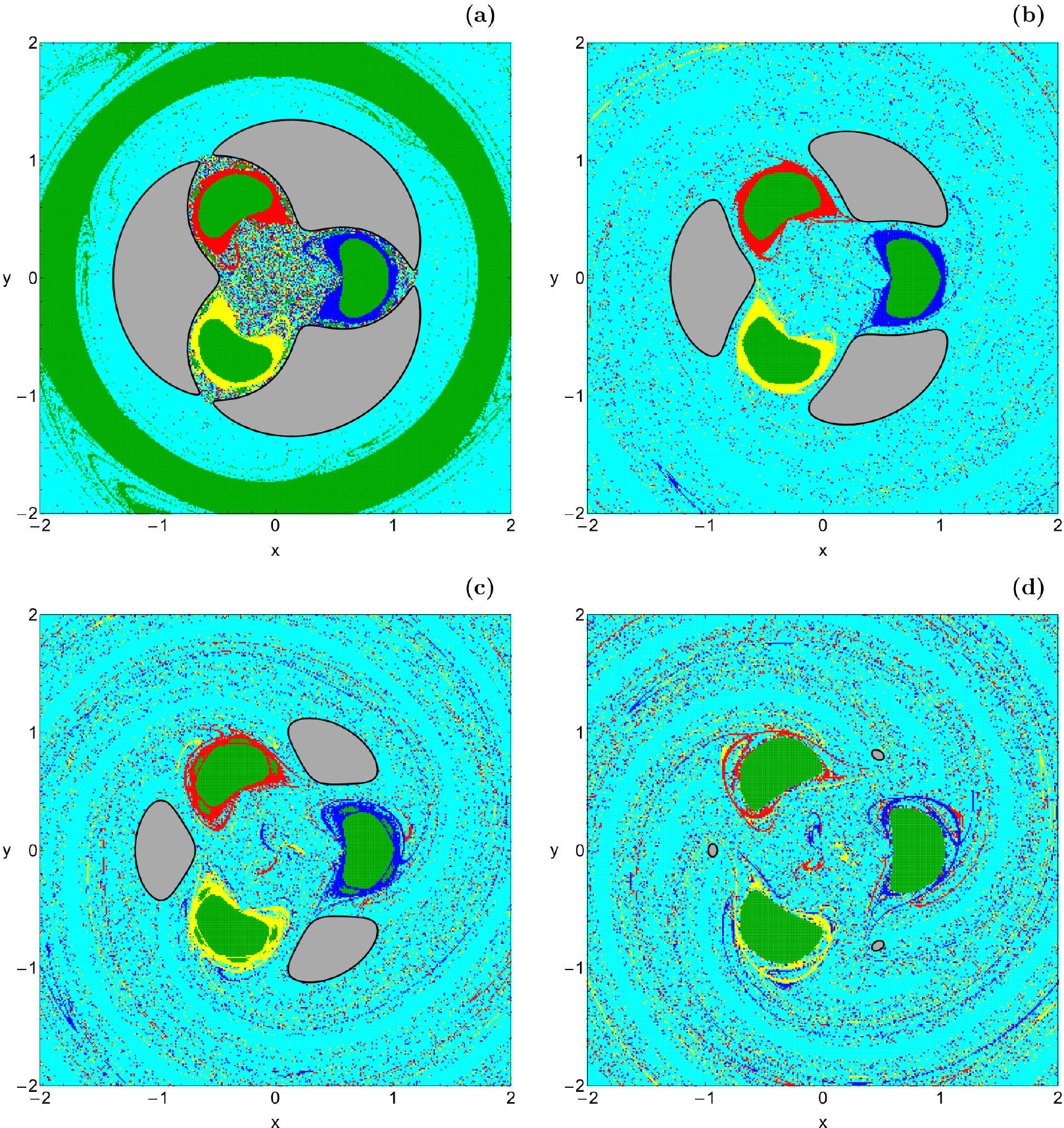}}
\caption{The orbital structure of the $\dot{\phi} < 0$ part of the surface of section $\dot{r} = 0$ when (a-upper left): $C = 3.355$; (b-upper right): $C = 3.22$; (c-lower left): $C = 3.09$; (d-lower right): $C = 2.95$. The color code is the same as in Fig. \ref{HR1}.}
\label{HR3}
\end{figure*}

\begin{figure*}[!tH]
\centering
\resizebox{\hsize}{!}{\includegraphics{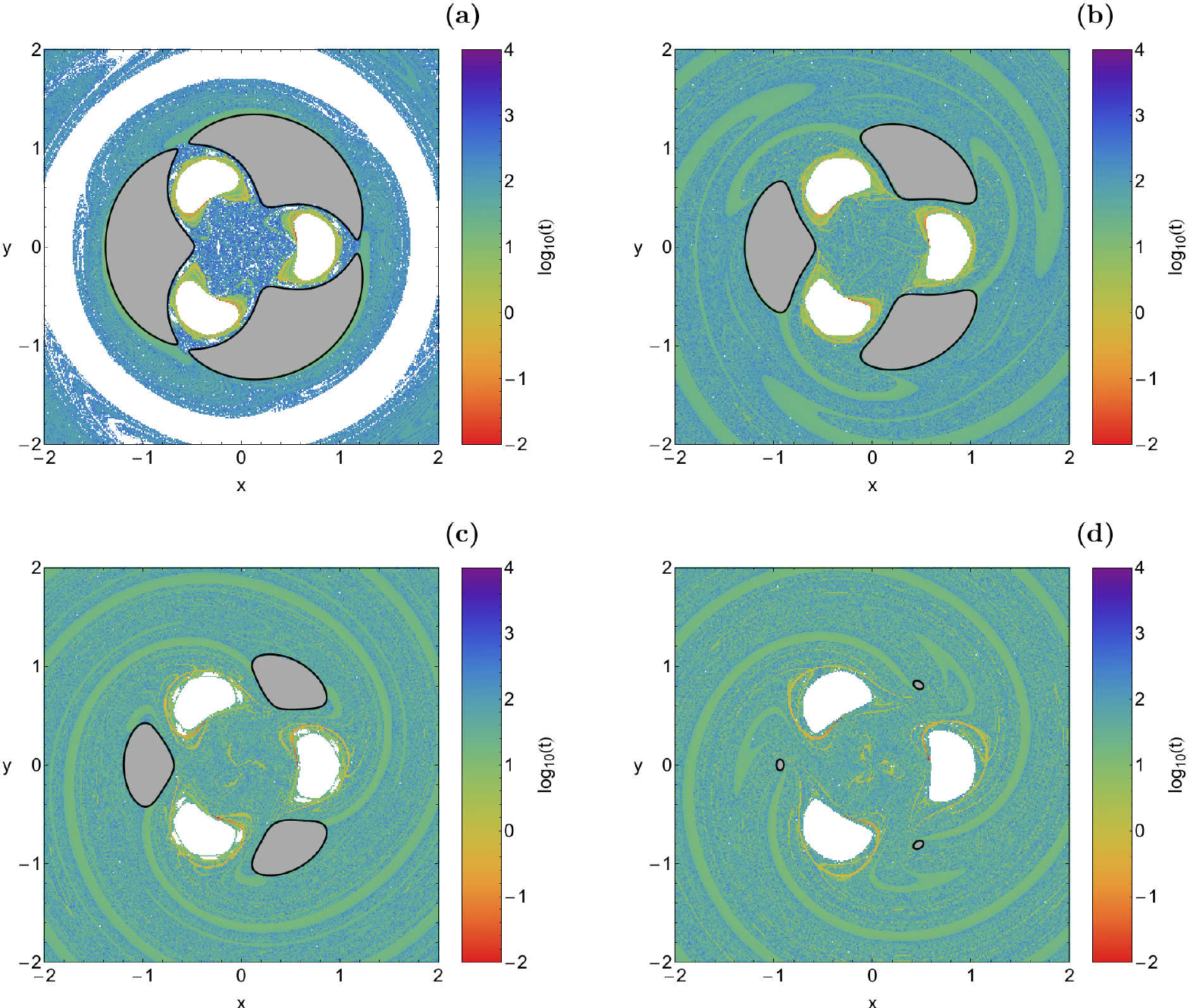}}
\caption{Distribution of the escape and collisional time of the orbits on the $\dot{\phi} < 0$ part of the surface of section $\dot{r} = 0$ for the values of the Jacobi constant of Fig. \ref{HR3}(a-d).}
\label{HR3t}
\end{figure*}

Case III concerns the third Hill's regions configurations where the fourth body with initial conditions in the interior region can escape from the system through one of the three symmetrical escape channels in the ZVCs. The following Fig. \ref{HR3}(a-d) shows the orbital structure of the configuration $(x,y)$ space through the OTD decompositions of the $\dot{\phi} < 0$ part of the surface of section $\dot{r} = 0$. In Fig. \ref{HR3}a where $C = 3.355$ we observe that the fractality of the central region is reduced, while initial conditions of escaping orbits are added in the mixture. The exterior region is still dominated by escaping orbits, however some lonely initial conditions of collisional orbits are present. For $C = 3.22$ it is seen in Fig. \ref{HR3}b that the outer stability annulus completely disappears. This behaviour was expected because we have seen in the previous energy cases that the stability of the annulus had started to reduce. As the value of the Jacobi constant decreases the fractality of the central interior region is reduced even further and for $C < 3.09$ one may observe in Fig. \ref{HR3}c that small collisional basins start to emerge near the center of the coordinates. At the same time, initial conditions of collisional orbits start to populate the exterior region, however we cannot speak of the formation of collisional basins yet. Looking carefully at Fig. \ref{HR3}(a-d) it becomes evident that the collisional basins around the stability islands of the primaries are reduced with decreasing $C$, however the size of the stability islands remains almost unperturbed.

In Fig. \ref{HR3t}(a-d) we illustrate the distribution of the escape and collisional time of orbits on the configuration $(x,y)$ space. We see that as soon as the stability annulus located in the exterior region disappears the escape rate of orbits is considerably smaller. Furthermore, we can easily distinguish the spiral escape basins in the exterior region in which the orbits escape from the system after about 10 time units of numerical integration.

\subsection{Case IV: $C < C_8$}
\label{sb4}

\begin{figure*}[!tH]
\centering
\resizebox{\hsize}{!}{\includegraphics{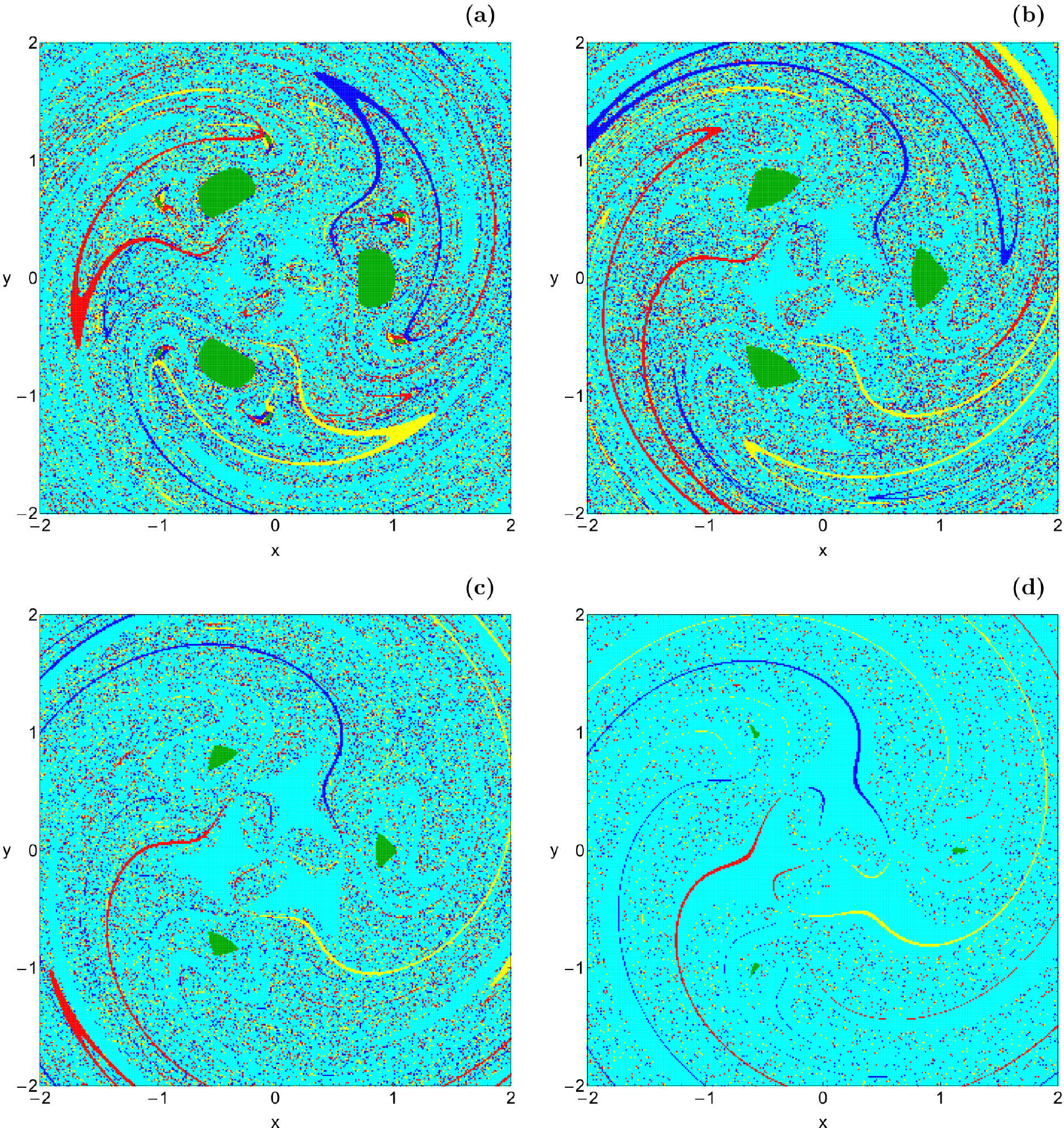}}
\caption{The orbital structure of the $\dot{\phi} < 0$ part of the surface of section $\dot{r} = 0$ when (a-upper left): $C = 2.50$; (b-upper right): $C = 2.30$; (c-lower left): $C = 2.10$; (d-lower right): $C = 1.90$. The color code is the same as in Fig. \ref{HR1}.}
\label{HR4}
\end{figure*}

\begin{figure*}[!tH]
\centering
\resizebox{\hsize}{!}{\includegraphics{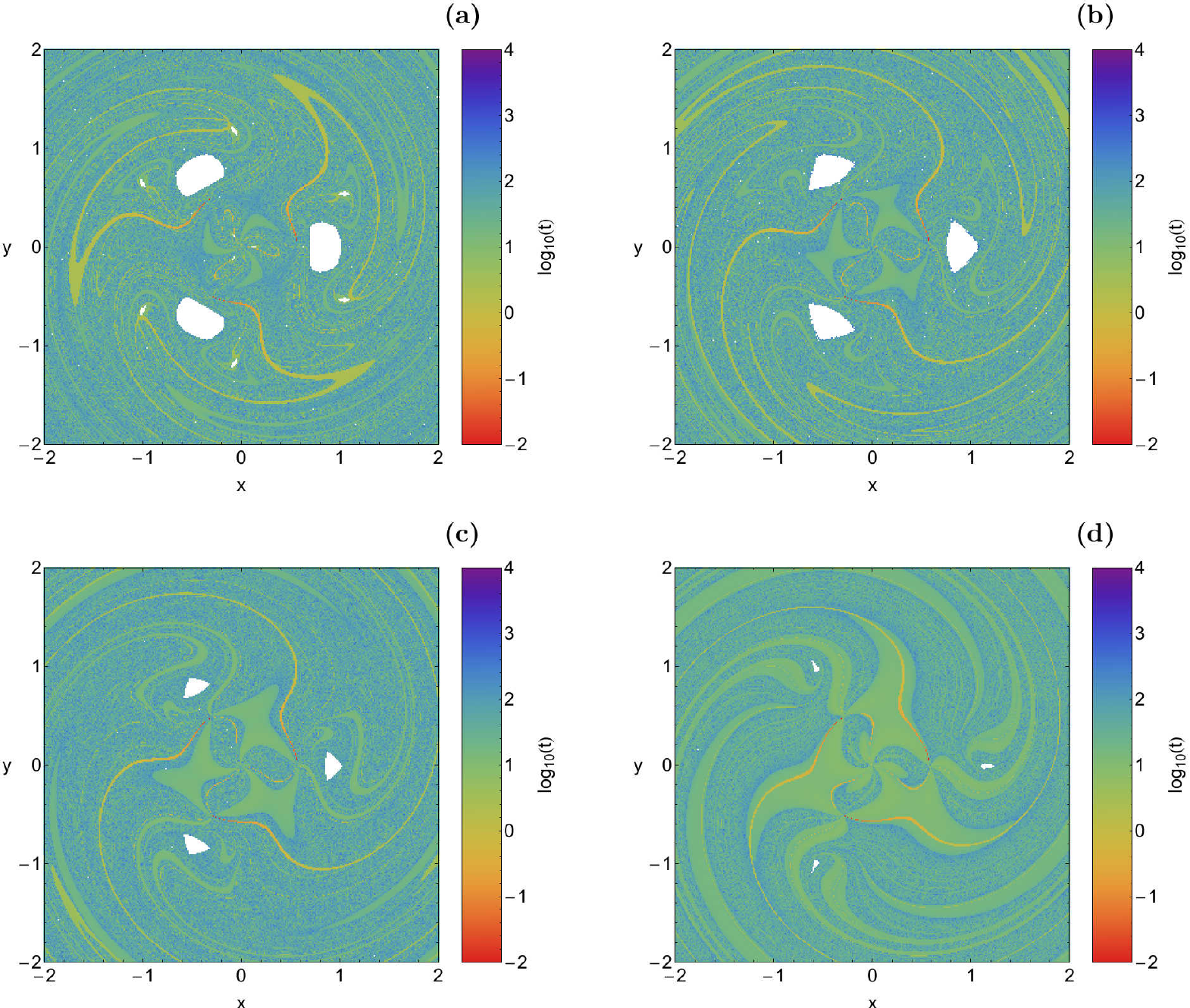}}
\caption{Distribution of the escape and collisional time of the orbits on the $\dot{\phi} < 0$ part of the surface of section $\dot{r} = 0$ for the values of the Jacobi constant of Fig. \ref{HR4}(a-d).}
\label{HR4t}
\end{figure*}

The last case under consideration involves the scenario when the test particle can freely travel all over the $(x,y)$-plane with no restrictions of forbidden regions. Once more, all the different aspects of the numerical approach remain exactly the same as in the two previously studied cases. Fig. \ref{HR4}(a-d) reveals the orbital structure of the configuration space through the OTD decompositions of the $\dot{\phi} < 0$ part of the surface of section $\dot{r} = 0$. In Fig. \ref{HR4}a, where $C = 2.50$, we observe that the collisional basins around the stability islands of the primaries are not present, while several well-formed collisional basins emerge in the exterior region. As the value of the Jacobi constant decreases, or in other words the total orbital energy increases, the most interesting phenomena which take place are the following: (i) the radii of the spiral structures of the collisional basins increase however the collisional basins become thinner; (ii) the size of the stability islands around the primaries is reduced and for $C < 1.90$ they disappear; (iii) the orbital content of the configuration $(x,y)$ space becomes poor since the portion of the initial conditions which correspond to escaping orbits heavily increases thus limiting the amount of collisional orbits.

The distribution of the escape and collisional times of orbits on the configuration space is depicted in Fig. \ref{HR4t}(a-d). One can see similar outcomes with that presented in the three previous subsections. At this point, we would like to emphasize that the basins of escape can be easily distinguished in Fig. \ref{HR4t}(a-d), being the regions with intermediate colors indicating fast escaping orbits. Indeed, our numerical computations suggest that orbits with initial conditions inside these basins need no more than 10 time units in order to escape from the system. Furthermore, the collisional basins are shown with reddish colors where the corresponding collisional time is less than one time unit of numerical integration. Indeed as the value of the Jacobi constant decreases several escape basins emerge and cover the vast majority of the configuration space. In particular, near the central region we can identify the presence of three escape basins which form a propeller shape.

The OTDs shown in Figs. \ref{HR1}, \ref{HR2}, \ref{HR3} and \ref{HR4} have both fractal and non-fractal (smooth) boundary regions which separate the three types of basins (escape, collisional and bounded). Such fractal basin boundaries is a common phenomenon in leaking Hamiltonian systems \cite{BGOB88,dML99,dMG02,STN02,ST03,TSPT04}. In the PERFBP system the leakages are defined by both escape and collision conditions thus resulting in four exit modes. However, due to the high complexity of the basin boundaries, it is very difficult, or even impossible, to predict in these regions whether the fourth body (e.g., a satellite, asteroid, planet etc) collides with one of the primary bodies or escapes from the dynamical system.

\subsection{An overview analysis}
\label{over}

\begin{figure}[!tH]
\centering
\includegraphics[width=\hsize]{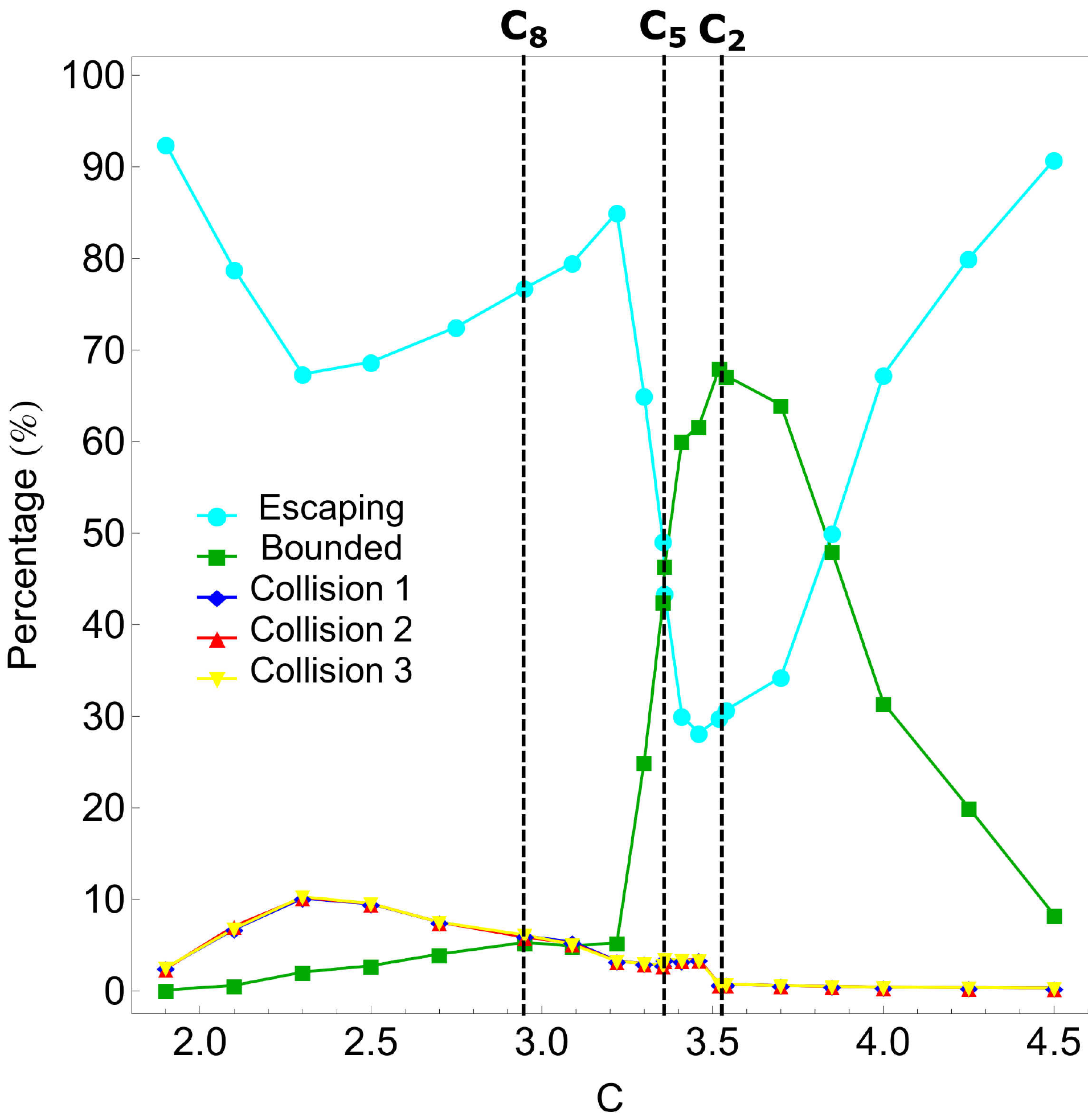}
\caption{Evolution of the percentages of the initial conditions of each considered basin as a function of the Jacobi constant. The vertical, dashed, black lines indicate the three critical values of $C$.}
\label{percs1}
\end{figure}

\begin{figure*}[!tH]
\centering
\resizebox{\hsize}{!}{\includegraphics{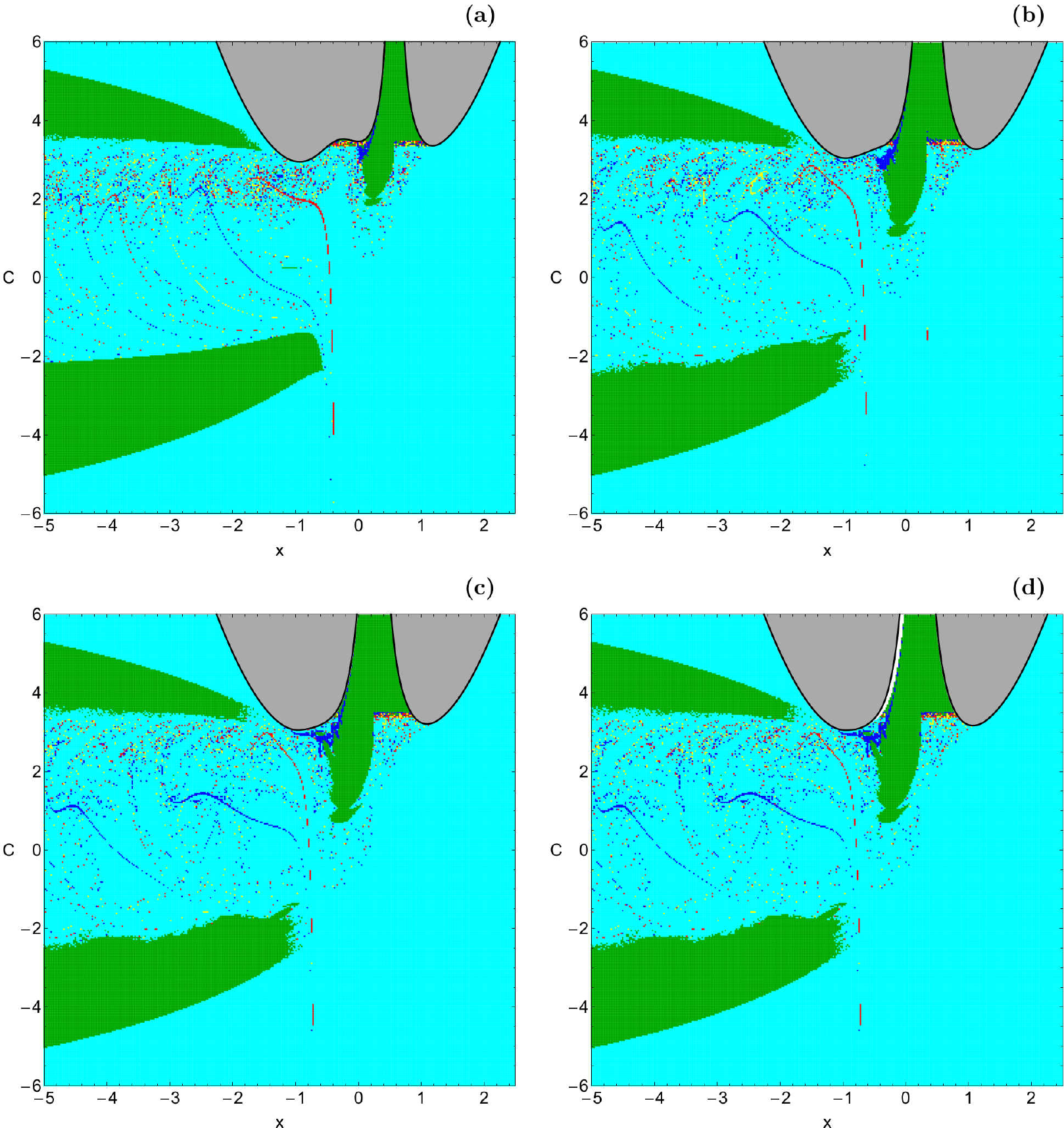}}
\caption{Orbital structure of the $(x,C)$ plane when (a-upper left): $\mu = \frac{1}{3}$; (b-upper right): $\mu = \frac{1}{5}$; (c-lower left): $\mu = \frac{1}{7}$; (d-lower right): $\mu = \frac{1}{9}$. The color code is the same as in Fig. \ref{HR1}.}
\label{xC}
\end{figure*}

\begin{figure*}[!tH]
\centering
\resizebox{\hsize}{!}{\includegraphics{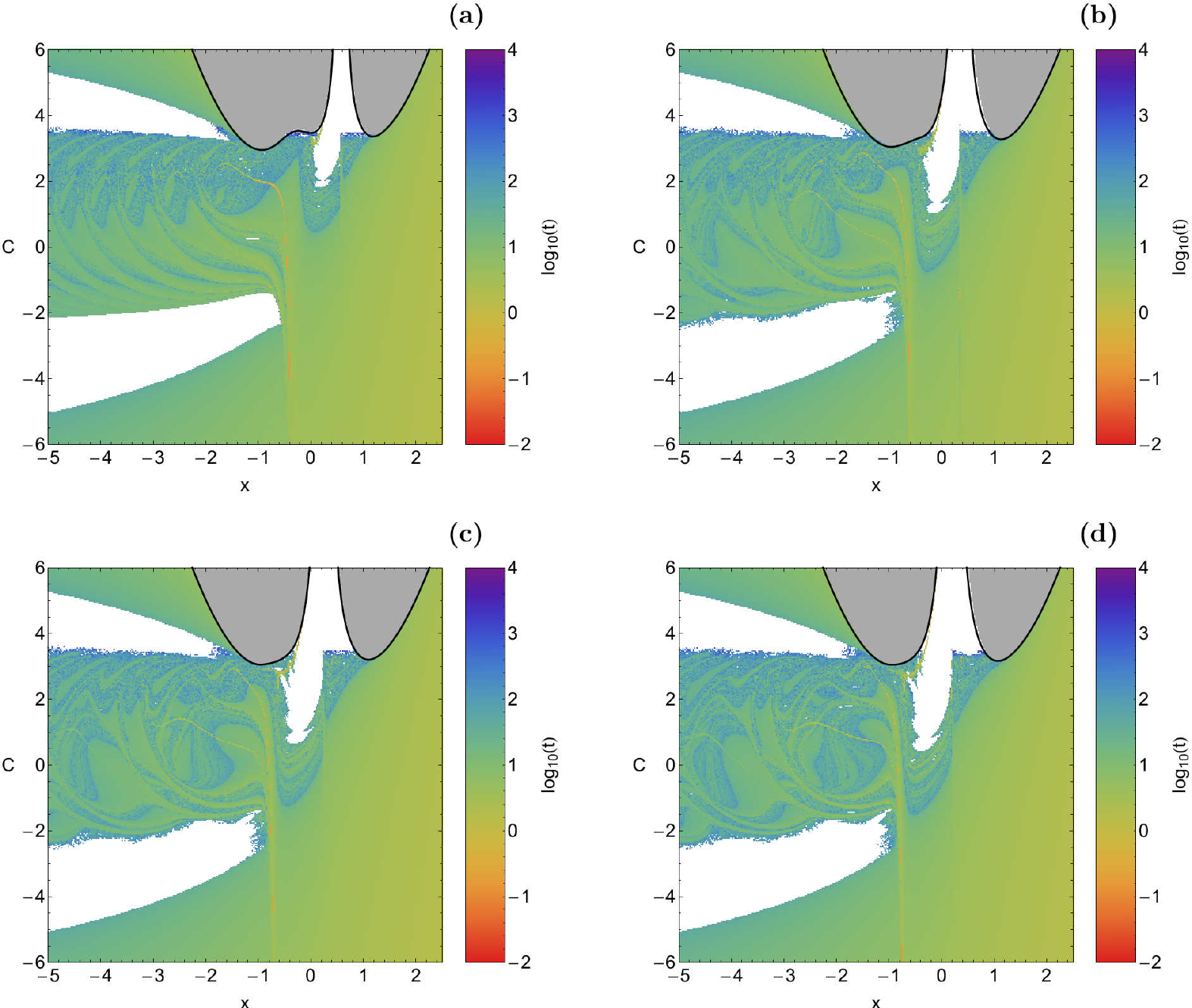}}
\caption{Distribution of the escape and collisional time of the orbits on the $(x,C)$ plane for the values of the radiation pressure factor of Fig. \ref{xC}(a-d).}
\label{xCt}
\end{figure*}

\begin{figure*}[!tH]
\centering
\resizebox{\hsize}{!}{\includegraphics{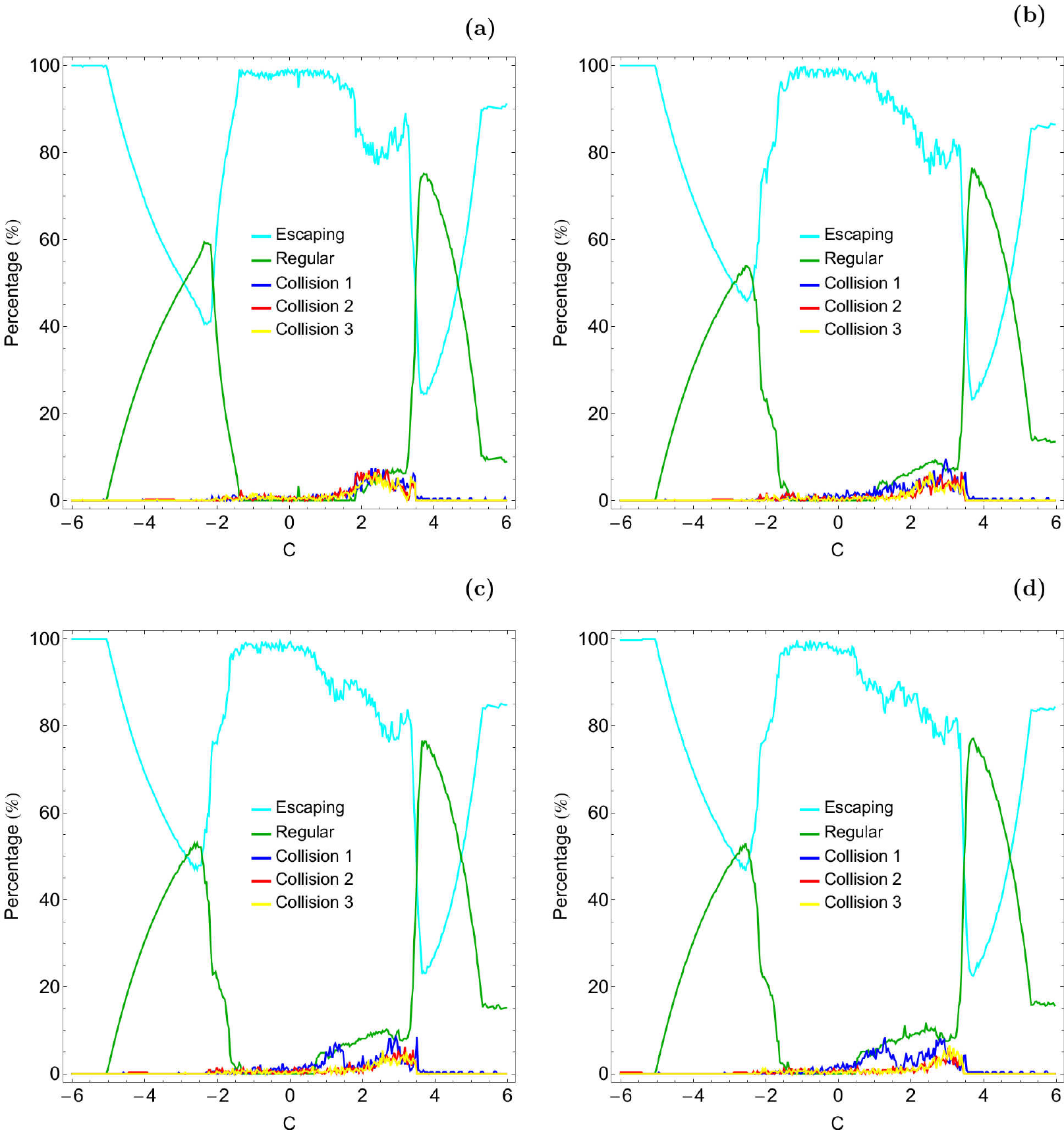}}
\caption{Evolution of the percentages of escaping, regular and collisional orbits on the $(x,C)$-plane as a function of the value of the Jacobi constant $C$. (a-upper left): $\mu = \frac{1}{3}$; (b-upper right): $\mu = \frac{1}{5}$; (c-lower left): $\mu = \frac{1}{7}$; (d-lower right): $\mu = \frac{1}{9}$.}
\label{pxC}
\end{figure*}

It would be very useful to monitor the evolution of the percentages of all types of orbits as a function of the Jacobi constant. Fig. \ref{percs1} shows such a diagram, where the vertical, dashed, black lines indicate the three critical values of the Jacobi constant $C$\footnote{Here we would like to point out that $C_1$ is not a critical value sice it does not affect the structure of the Hill's regions configurations.}. We decided not to include in the diagram the rate of trapped chaotic orbits mainly because its value is always extremely low (lower than 1\%). It is observed that for $C < 3.3$ escaping orbits is the most populated family and especially for $C < 2.9$ they occupy more than 90\% of the configuration $(x,y)$ space. However, for $C > 3.3$ the rate of escaping orbits exhibits a dramatically decrease until $C = 3.45$. For $C > 3.45$ the portion of escaping orbits increases and for $C > 4.5$ they cover, once again, more than 90\% of the $(x,y)$-plane. Therefore we may say that escaping orbits dominate for both very low and very high values of the Jacobi constant. The percentage of bounded regular orbits increases almost linearly in the interval $C \in [1.9, 3.2]$, while for $C > 3.2$ we observe a sudden increase until $C = C_2$, where regular bounded orbits cover about 70\% of the configuration space. For $C > C_2$ the same rate drops steadily and at the highest value of the Jacobi constant studied $(C = 4.5)$, regular orbits occupy about one tenth of the $(x,y)$-plane. The three percentages of collisional orbits are exactly the same due to the symmetry of the PERFBP. In most cases (i.e., when $C \in [3.5, 4.5]$) the rates of collisional orbits are very low (lower than 5\%). The highest rate of the collisional orbits was reported for $C = 2.3$ where all types of collisional orbits cover about 30\% of the configuration space.

The color-coded OTDs in the configuration $(x,y)$ space provide sufficient information on the phase space mixing however, for only a fixed value of the Jacobi constant (or the total orbital energy) and also for orbits that traverse the surface of section retrogradely. H\'{e}non \cite{H69}, introduced a new type of plane which can provide information not only about stability and chaotic regions but also about areas of bounded and unbounded motion using the section $y = \dot{x} = 0$, $\dot{y} > 0$ (see also \cite{BBS08}). In other words, all the initial conditions of the orbits of the test particles are launched from the $x$-axis with $x = x_0$, parallel to the $y$-axis $(y = 0)$. Consequently, in contrast to the previously discussed types of planes, only orbits with pericenters on the $x$-axis are included and therefore, the value of the Jacobi constant $C$ can now be used as an ordinate. In this way, we can monitor how the energy influences the overall orbital structure of our dynamical system using a continuous spectrum of Jacobi constants rather than few discrete values. In Fig. \ref{xC}(a-d) we present the orbital structure of the $(x,C)$ plane for four values of the mass parameter $\mu$ when $C \in [-6,6]$, while in Fig. \ref{xCt}(a-d) the distribution of the corresponding escape and collision times of the orbits is depicted. The black solid line in Fig. \ref{xC}(a-d) is the limiting curve which distinguishes between regions of allowed and forbidden motion and is defined as
\begin{equation}
f_L(x,C) = 2\Omega(x,y = 0) = C.
\label{zvc}
\end{equation}

We observe the presence of several stability regions. Being more precise, between the forbidden regions we identify a stability island corresponding to both direct (counterclockwise) and retrograde (clockwise) quasi-periodic orbits around primary 1. It is seen that a large portion of the exterior region is covered by initial conditions of escaping orbits however, at the left-hand side of the $(x,C)$-plane two stability islands of regular orbits that circulate around all three primaries are observed. We also see that between these two stability islands there are several thin basins corresponding to collisional orbits to all primary bodies. At the right-hand side of the $(x,C)$-plane on the other hand, a vast basin of escape is present, while there is no indication of collisional motion whatsoever. It should be pointed out that in the blow-ups of the diagrams several additional very small islands of stability have been identified\footnote{An infinite number of regions of (stable) quasi-periodic (or small scale chaotic) motion is expected from classical chaos theory.}.

It is evident that the particular value of the mass ratio $\mu$ does not really change the qualitative nature of the orbital structure of the PERFBP. Looking carefully at Fig. \ref{xC}(a-d) however, we can distinguish some minor changes with the increase of the mass ratio: (i) The size of the stability island between the forbidden regions increases; (ii) The orbital content in the exterior region between the two stability islands becomes increasingly poor, as the escape domains take over; (iii) For $\mu = \frac{1}{9}$ at the boundaries between the stability island and the left forbidden region we observe the presence of a thin collisional basin to primary 3 which is absent in the other three cases.

Finally, it would be very informative to monitor the evolution of the percentages of the different types of orbits as a function of the Jacobi constant $C$ for the $(x,C)$-planes shown in Figs. \ref{xC}(a-d). Our results are presented in Figs. \ref{pxC}(a-d). We see that in all four cases the percentages display similar patterns, so we are going to explain only the first case shown in Fig. \ref{pxC}a where $\mu = \frac{1}{3}$. For $C < -5$ escaping orbits cover all the available space however their rate gradually reduces until about 40\% for $C = -2.5$. For $C > -2.5$ it suddenly increases and in the interval $[-1.5, 3]$ escaping orbits dominate with rates above 80\%. The evolution of the percentage of regular bounded orbits displays an exact opposite evolution with respect to the pattern of escaping orbits. In particular, in the interval $[-1.5, 3]$ the percentage of regular orbits fluctuate at very low values below 10\%, while for $C < -2$ and $C > 3$ on the other hand, two peaks are observed at about 60\% and 75\%, respectively. The percentages of collisional orbits to primaries 1, 2 and 3 have a monotone behaviour with almost zero values for $C < 2$ and $C > 3.5$, while in the interval $[2, 3.5]$ their rates fluctuate at extremely low values below 5\%. Thus we may argue that the most interesting interval of values of the Jacobi constant $C$ is the interval $[-2, 3.5]$. Therefore one may reasonably conclude that the mass parameter practically does not drastically affect the actual percentages of the different types of orbits. All it does is to change the minima and the maxima of the different percentages.

\section{Discussion and conclusions}
\label{conc}

The scope of this research work was to reveal the orbital structure of the planar equilateral restricted four-body problem (PERFBP) with three equal masses. After conducting an extensive and thorough numerical investigation we managed to distinguish between bounded, escaping and collisional orbits and we also located the basins of escape and collision, finding also correlations with the corresponding escape and collision times. Our numerical results strongly suggest that the motion of the test particle under the gravitational field of the three primaries is very complicated. To our knowledge, this is the first detailed and systematic numerical analysis on the escape and collisional dynamics of the PERFBP and this is exactly the novelty and the contribution of the current work.

For several values of the Jacobi constant in the all possible Hill's regions configurations we defined dense uniform grids of $1024 \times 1024$ initial conditions regularly distributed on the $\dot{\phi} < 0$ part of the configuration $(x,y)$ plane inside the area allowed by the value of the Jacobi constant (or in other words by the value of the total orbital energy). All orbits were launched with initial conditions inside the scattering region, which in our case was a square grid with $-2\leq x,y \leq 2$. For the numerical integration of the orbits in each type of grid, we needed about between 8 hours and 5 days of CPU time on a Pentium Dual-Core 2.2 GHz PC, depending on the escape and collisional rates of orbits in each case. For each initial condition, the maximum time of the numerical integration was set to be equal to $10^4$ time units however, when a particle escaped or collided with one of the two primaries the numerical integration was effectively ended and proceeded to the next available initial condition.

We provide quantitative information regarding the escape and collisional dynamics in the PERFBP. The main outcomes of our numerical research can be summarized as follows:
\begin{enumerate}
 \item In the special case of the PERFBP where all three primary bodies have equal masses $(\mu = \frac{1}{3})$ there are ten unstable equilibrium points. Due to the equality of the masses of the primaries the system admits a symmetry and the equilibrium points lie on the $(x,y)$-plane symmetrically to three axes of symmetry.
 \item In all examined cases, areas of bounded motion and regions of initial conditions leading to escape or collision, were found to exist in the configuration space. There are two main\footnote{It should be emphasized and clarified that apart form the regular orbits that circulate around one or around all three primaries there are also other types of regular orbits. In particular, there are many families of periodic orbits around the equilibrium points (the Lyapunov periodic orbits is a characteristic example of such orbits). Our numerical calculations however strongly suggest that in the PERFBP the majority of bounded regular motion corresponds to regular orbits circulating around the primaries.} types of regular bounded orbits: (i) orbits that circulate around one of the primaries and (ii) orbits that circulate around all three primary bodied.
 \item In the first Hill's regions configuration where the transitions channels are still closed we found for every primary inside the corresponding stability islands collisional basins, while the exterior region is mostly occupied by escaping orbits. In this energy case a stability annulus is always present in the exterior region.
 \item In the second Hill's regions configurations when the transition channels open we detected the existence of trapped delocalized chaotic orbits. The central interior region appears to be highly fractal, while the stability annulus in the exterior region starts to destabilizes as the value of of the Jacobi constant decreases.
 \item It was observed that the stability ring in the exterior region completely disappears in the third Hill's regions configurations. At the same time the exterior region fills with initial conditions of orbits leading to collision with the primaries. On the other hand, the collisional basins around the stability islands start to fade as we proceed to lower values of $C$, or in other words higher values of the total orbital energy.
 \item Our calculations reveal that in the last Hill's regions configurations the collisional basins extend to the entire configuration space. However their size is constantly reduced with decreasing value of the Jacobi constant, while at the same time the basins of escape take over the $(x,y)$-plane.
 \item We presented numerical evidence through the construction of the $(x,C)$-planes that the value of the mass parameter $\mu$ does not practically influence the orbital content of the dynamical system. The value of $\mu$ affects only the number as well as the location of the equilibrium points of the PERFBP.
\end{enumerate}

Judging by the detailed and novel outcomes we may say that our task has been successfully completed. We hope that the present numerical analysis and the corresponding results to be useful in the field of escape dynamics in the PERFBP. The results as well as the conclusions of the present research are considered, as an initial effort and also as a promising step in the task of understanding the escape mechanism of orbits in this interesting dynamical system. Taking into account that our outcomes are encouraging, it is in our future plans to properly modify our dynamical model in order to expand our investigation into three dimensions and explore the entire six-dimensional phase. Furthermore, it would be of particular interest to add perturbations (i.e., oblateness, radiation pressure, etc) in the model-potential and try to understand their influence on the character of orbits.



\end{document}